\documentclass[11pt,a4paper]{article}
\usepackage{graphicx}
\begin{document}
\title{Generalized Statistics and High $T_c$ Superconductivity}
\author{H. Uys \thanks{E-mail:    huys@maple.up.ac.za}, H. G. Miller \thanks{E-Mail:   hmiller@maple.up.ac.za}\\ \textit{\small Department of Physics, University of Pretoria, Pretoria 0002, South Africa} \and F. C. Khanna \\ \textit{\small Theoretical Physics Institute, Department of Physics, University of Alberta,} \\ \textit{\small Edmonton, Alberta, Canada T6G~2J1}  \\ \textit{\small TRIUMF, 4004 Wesbrook Mall, Vancouver, British Columbia, Canada V6T~2A3 }
}
\date{16 May 2001}

\maketitle

\begin{abstract}
 Introducing  the generalized, non-extensive statistics  proposed by Tsallis \cite{tsallis88}, into the standard s-wave pairing BCS theory of superconductivity in 2D 
yields a reasonable description of many of the main properties of high temperature superconductors, provided some allowance is made for non-phonon mediated interactions.

\end{abstract}

    The discovery of superconductivity in the copper oxides in 1986 \cite{bednorz86} and the subsequent race for even higher critical 
temperatures (125K by 1993 \cite{putilin93,schilling93}) raised hopes for the application of  superconducting phenomenon  at operating temperatures approaching room temperature.
  The inability of the Bardeen-Cooper-Schrieffer (BCS) model  \cite{BCS} to describe 
superconductivity in these materials satisfactorily, appears to indicate that we are dealing with a  completely different class of superconductors.  Various 
theoretical models have  been proposed to explain this phenomenom, ranging from d-wave superconductivity \cite{annett95,muzikar97} to models 
incorporating non-phononic coupling mechanisms \cite{alexandrov81,varma89,anderson87,little64}.  Different degrees of success have been achieved in explaining specific aspects of these high-$T_c$ materials,  but no inclusive model exists.

    A number of characteristics must be incorporated in any such  model.   Certainly the main common denominator in all  high-$T_c$ 
materials is the large anisotropy in the crystal structure, resulting in conduction electron states in the CuO$_2$ planes being very 
nearly two-dimensional in character.   It is therefore reasonable to assume that this is essential for the high critical 
temperatures  and that arguably the 'ideal' high-$T_c$ superconductor is purely two-dimensional in character.  It is, however, well 
known that in BCS,  a pure 2D model presents problems as the phase loses its coherence due to fluctuations 
\cite{hohenberg67}.   This calls for at least a quasi-2D model in all realistic cases.    Although the results of measurements on the 
superconducting  gap are somewhat varied, there seems to be convergence towards a ratio of  $\frac{2\Delta_0}{k_B T_c}$ equal to 
somewhere between 6 and 8 
which is not consistent with the universal value of $\sim 3.5$ obtained for normal BCS superconductors.  This is particularly so in the case of tunneling measurements \cite{briceno89,edwards92,zhang93,jeong94},
 where ratios as high as $8.9$ have been measured in La$_{1.85}$Sr$_{0.15}$CuO$_4$ \cite{ekino96}, but  appears to  contradict optical 
measurements \cite{bonn87,degiorgi87,collins87} which seem to yield results closer to the 
BCS ratio for weak coupled superconductors.
 In spite of the apparent BCS behaviour of the tunneling currents, the violation of  universality  is difficult to understand. 
Thirdly,  experimental data suggests that the electronic contribution to the  specific heat in 
cuprates does not exhibit the exponential form of normal weak coupled superconductors. It is rather of the linear form  $\gamma$T, 
where $\gamma$ is of the order of a few mJ/mol$\cdot$K$^2$.   Magnetic properties of the high-$T_c$ superconductors also differ appreciably from their normal 
counterparts,  most notably in their very high upper critical fields.  These fields are extrapolated to be of the order of tens or 
hundreds of Teslas and are too large to be attained with other present technologies. Furthermore  it should also be noted  that all high-$T_c$ 
superconductors are type II  superconductors.  One of the strongest arguments for the phonon coupling scheme of the 
BCS model is the existence of an isotope effect.  In cuprate materials, where there is a strong doping dependence of the isotope effect, this requires some additional consideration.  In optimally doped materials the effect is often strongly suppressed ($T_c$ may scale as a power $\alpha = 0.1$ of the mass), whereas in overdoped or underdoped materials it might be more prevalent, with the scaling 
in some cases even exceeding $\alpha = 0.5$ \cite{waldram96}.

Recently in statistical mechanics there has been interest in  a generalization of the Boltzmann-Gibbs (BG) 
statistics to a non-extensive form proposed by Tsallis\cite{tsallis88} in  which the former is recovered in an appropriate limit.  This formalism 
has had considerable success in providing an appropriate mathematical framework for dealing with physical systems with long-range 
interactions.  This is markedly so in the so-called stellar "polytropes" \cite{plastino93} where the usual Boltzman distribution 
functions yield unphysical results.  A variety of other applications have also been considered along with the generalization of many 
well-known theorems and principles, see for example the references in \cite{plastino94}.  

   In this paper we propose and motivate the possibility that the cuprate oxide materials  have an inherent underlying 
non-BG like character responsible for their high-critical temperatures and other unusual properties.   We suggest that the s-wave BCS 
model is essentially correct in employing an effective weak coupling Hamiltonian that includes the kinetic energy of free electrons along with a  constant 
attractive potential between electrons of equal momentum and opposite spin \cite{BCS}.   The BCS ansatz in a purely 2D form is used for the ground state,
 with a generalized form of the Fermi-distribution function \cite{fevzi95} appearing at finite temperatures.  We show elsewhere that such a 2D model is justified as the fluctuation-dissipation theorem is no longer valid in the Tsallis formalism \cite{miller01}. The assumption that the coupling mechanism is purely phonon mediated, needs to be 
modified to obtain reasonable fits to the experimental data, particularly in the case of very high temperature superconductng materials.

  The relevant question is, of course, why one would resort to changing the entropic measure in a high-$T_c$ superconductor, and if 
one does, why the Tsallis formulation should be appropriate.  A very compelling argument can be found from a consideration of the electronic 
specific heat.  The more exotic coupling methods mentioned above, certainly may have the effect of removing problems indirectly related 
to the BCS-Hamiltonian such as the lattice instabilities predicted by Migdal \cite{migdal58}.  The experimental evidence of an energy gap, however, requires an energy spectrum for the total energy of excitations, of the form
\begin{equation}
\label{spectrum}
E_k^2 = \varepsilon_k^2 + \Delta^2, \label{qpe}
\end{equation}
where  $\varepsilon$  is the excitation energy and $\Delta$ the energy gap, independent of any proposed model.  
Assuming a Fermi distribution, the electronic specific heat capacity ($C$) at temperatures $kT \ll \Delta$,  can shown to  have the following form \cite{lifshitzvol9}  
\begin{equation}
\label{expC}
C \;\; \sim \;\; \frac{e^{-\frac{\Delta_0}{kT}}}{T^\frac{3}{2}}
\end{equation}
One is therefore obliged to contend with an exponential form of the heat capacity in any model that relies
on the Fermi distribution function used in BG statistical mechanics.  One way to circumvent this is to introduce a 
different distribution function.

   It should be remembered that the most basic interaction in an electron gas is that of the Coulomb repulsion. This interaction is of 
infinite range and may also introduce correlation effects. One example of this is the condensation of a true electron gas, of low enough
density, into a \textit{Wigner lattice} \cite{wigner34}. Very early attempts at explaining superconductivity, quite intuitively, but unsuccesfully used 
the Coulomb potential as a starting point \cite{heisen47,koppe57}.  More recently models like those of Hubbard \cite{hubbard63}, describing 
contributions due to repulsion between electrons sharing an atomic orbital, have been applied to superconductivty.   The condensation of electrons into its 
superfluid state is ultimately the result of an effective, attractive interaction.  The extent of the effect of Coulomb repulsion might
not be obvious and in spite of the fact that we do not explicitly take it into account in the effective interaction, 
 we argue that  it is precisely the effects of this long-range interaction which may render the system more suitable for description by 
generalized statistics.

The  generalized entropy postulated by Tsallis \cite{tsallis88} takes the form:

\begin{equation}
\label{qentropy}
S_q = k \frac{\sum^w_{i=1} \{ p_i - p^q_i \}}{q-1}      \;\;\;\;       (q \in \Re)
\end{equation}
where w is the total number of microstates in the system and p is the associated probabilities. It has been  shown that this entropy obeys
the usual properties of concavity, equiprobability, positivity and irreversibility and Shannon additivity and preserves the 
Legendre transformation  stucture of thermodynamics \cite{curado91}.  It is also straight forward to verify that the usual BG entropy is recovered in 
the limit $q\rightarrow 1$.  Associated with equation (\ref{qentropy}) is the generalized Fermi distribution  given by \cite{fevzi95}

 \begin{equation}
\label{qfermi}
f_{q} = \frac{1}{[1+\beta(q-1)\epsilon_k]^\frac{1}{q-1} +1}.
\end{equation}
Once again the usual Fermi distribution of BG statistics is recovered in the limit $q\rightarrow 1$. It is this generalized 
distribution, we will henceforth assume, that  excited  independent quasi-particles  obey at finite temperature.  
In 2D
the BCS gap equation for weak coupled superconductors at finite temperature \cite{BCS} is given as 
\begin{equation}
\label{gapeq}
\Delta_p = -\sum_k g_{kp} \frac{\Delta_k}{2E_k} (1- f_{k\uparrow} - f_{k\downarrow})
\end{equation}
where $g$ is the coupling constant, $E_k$ is the quasi-particle energy given by equation (\ref{qpe}) and  $f_{k\sigma}$ is the  Fermi distribution function.
Replacing  $f_{k\sigma}$ by equation (\ref{qfermi}) and  the sum by an integral over the density of states  we obtain in 2D a generalized form for the gap equation:

\begin{equation}
\label{qgap}
\frac{1}{N(0)g} = \int \frac  {d \varepsilon}{\sqrt{\varepsilon^2 + \Delta^2}}\frac {{[1+\beta(q-1)\sqrt{\varepsilon^2 + \Delta^2}
]^\frac{1}{q-1} } - 1} {{[1+\beta(q-1)\sqrt{\varepsilon^2 + \Delta^2}
]^\frac{1}{q-1} } + 1}
\end{equation}
where $N(0)$ represents the density of states.
For q=1 in 3D, one recovers the gap quation in the standard BCS theory and, of course, a description  of normal
superconductors in the weak coupling limit.
 As required, equation (\ref{qgap}) is independent of the distribution function and at zero 
temperature reduces to
\begin{equation}
\label{0Tgap}
\frac{1}{N(0)g} = \int\frac{d \varepsilon}{(\Delta_0^2 + \varepsilon^2)^\frac{1}{2}}
\end{equation}
where $\Delta_0$ is the zero temperature gap. 

Let us, as a starting point, 
choose as cutoff to the gap equation the Debye frequency $\hbar\omega_D$.   The  analytical solution to equation (\ref{0Tgap}) is 

\begin{equation}
\Delta_0 = (\hbar\omega_D + \sqrt{\hbar\omega_D^2 + \Delta_0^2}) e^\frac {-1}{N(0)g}.
\end{equation}
Using experimental values for the gap and Debye frequency, one can solve for $N(0)g$ at zero temperature. The problem then reduces to 
finding that $q$  in the gap equation, which yields a vanishing gap at the critical temperature.  With $q$ thus fixed, the 
temperature dependence of the gap can be determined.

Consider again the gap in equation (\ref{gapeq}) at $T_c$, independent of the particular choice of statistics.   A change in the integration variable $\varepsilon \rightarrow \epsilon k_B T_c$ 
may always be made such that the integration limits are from $0$ to $\frac{\theta_D}{T_c}$.  Integrating by parts yields:

\begin{equation}
\frac{1}{N(0)g} = (1-2f(\frac{\theta_D}{T_c}))\ln{\frac{\theta_D}{T_c}} - \int_0^\frac{\theta_D}{T_c} (\epsilon - 2F(\epsilon))\ln{\epsilon} \hspace{3 mm} d\epsilon
\end{equation}
where F($\epsilon$) is the indefinite integral of f($\epsilon$).
Taking the integral on the right to the left hand side and   dividing by $(1-2f(\frac{\theta_D}{T_c}))$ yields some number $\varphi$ which depends, of course, on the choice of $f(\epsilon)$ . Exponentiating both sides yields:

\begin{equation}
\label{iso}
T_c=\theta_D e^{\varphi}
\end{equation}
which clearly demonstrates that the isotope effect is preserved in the BCS formulation irrespective of the particular form  of statistical mechanics employed.

  It is at this stage appropriate to note that despite reasonable physical arguments in the 
case of normal superconductors, the choice of the Debye frequency as a cutoff is mathematically quite arbitrary.  The temperature 
dependent solution to the gap equation in fact converges to a well defined value for any cutoff of sufficient magnitude.  It is only when
the cutoff is of a magnitude of the order of the gap itself, that deviations occur.  In normal superconductors the Debye frequency is 
greater than the gap by $\sim 10^2$ and thus adequate.  This is, however, certainly not the case in high-$T_c$ superconductors where the gap
at zero temperature may even be larger than the Debye frequency, e.g. in Tl$_2$Ba$_2$Ca$_2$Cu$_3$O$_{10}$ \cite{hoffmann94}  (see \cite{ginsberg90} for Debye frequencies of other high $T_c$ superconductors).
Abandoning the Debye frequency as a cutoff is tantamount to acknowledging that the electron-phonon 
interaction is not the only interaction involved in forming the condensed state.  Changing the cutoff will, of course, influence the 
isotope effect.  It might be  argued that the observed suppression in the isotope effect may be a consequence of this.

It is interesting to note that the convergence of the gap at greater cutoffs, is accompanied by a simultaneous 
convergence of $q$.  Consider a transformation of the gap in equation (\ref{qgap}) via the substitution $\varepsilon = \frac{k_B T_c 
\varepsilon^\prime}{2}$.  Let us also define the cutoff in terms of multiples of the energy gap,
\textit{e.g.} $n\Delta_0$.  Then for
a given value of the gap to critical temperature ratio, $\frac{2\Delta_0}{k_B T_c}= m$, one can show that the gap equation reduces to a
form dependent only on the ratio $\frac{2\Delta_0}{k_B T_c}$, the number $n$ and $q$, and not on either the gap or critical temperature 
directly.

\begin{equation}
\label{dimlsgap}
\frac{1}{N(0)g} = \int_0^{m n} \frac  {d \varepsilon^\prime}{\varepsilon^\prime} \frac{[1+(q-1)\varepsilon^\prime/2
]^\frac{1}{q-1}  - 1} {[1+(q-1)\varepsilon^\prime/2]^\frac{1}{q-1}  + 1}
\end{equation}
Thus, if for a specific ratio of $\Delta_0$ to $T_c$, the temperature dependent gap is convergent independent of 
cutoff, then $q$ must have a uniquely defined value.   It therefore seems appropriate to define the cutoff in terms of the convergence of  $q$  as this can be specified independent of the critical temperature pertaining to a specific material.  

\begin{figure}[h]
\centering
\begin{center}
\includegraphics[scale=0.5,angle=270]{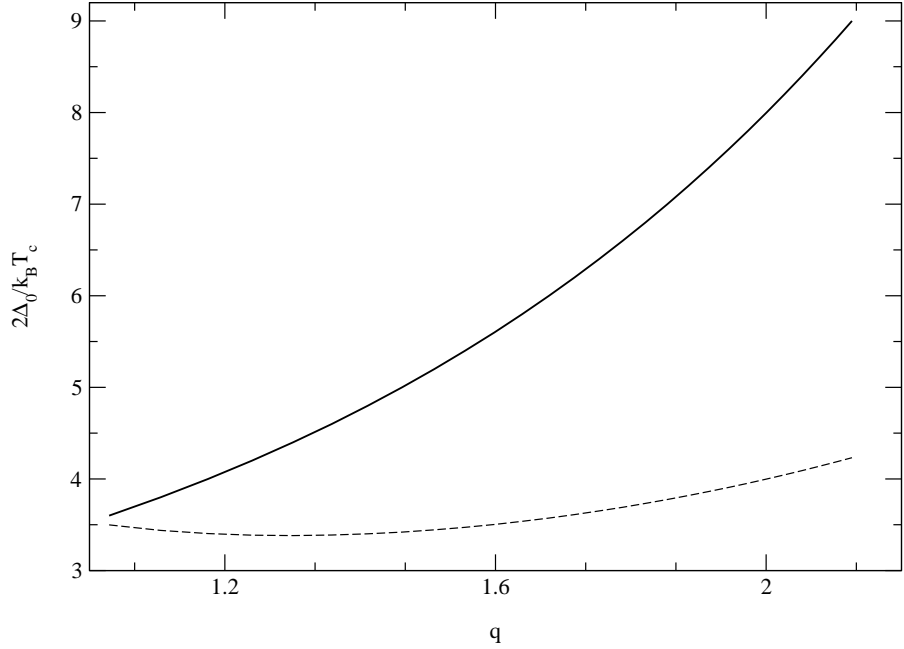}
\end{center}
\label{fig1}\caption{The ratio $\frac{2\Delta_0}{k_B T_c}$ vs. q is given by the solid curve. For comparison the dashed curve representing
  $\frac{2\Delta_0}{qk_B T_c}$ vs. q is included.
 }
\end{figure}

A relevant question now is whether a generalization of the BCS universality condition of $\frac{2\Delta_0}{k_BT_c}\sim 3.52$ exists for the generalized statistics of Tsallis. Clearly the generalization must be $q$ dependent because of the dependence of $T_c$ on $q$ and reduce to the BCS universality condition for $q$=1. In 
Fig.1 a graph  of $\frac{2\Delta_0}{q k_BT_c} $ versus $q$ is given. This ratio does not deviate appreciably from 3.5 which suggests 
the following generalization of the universality condition
\begin{equation}
\frac{2\Delta_0}{q k_BT_c}\sim 3.52
\end{equation}
A more detailed analysis might lead to replacing $q$ by some function of $q$ which should be $\sim q$.

\begin{figure}[h]
\centering
\begin{center}
\includegraphics[scale=0.5,angle=270]{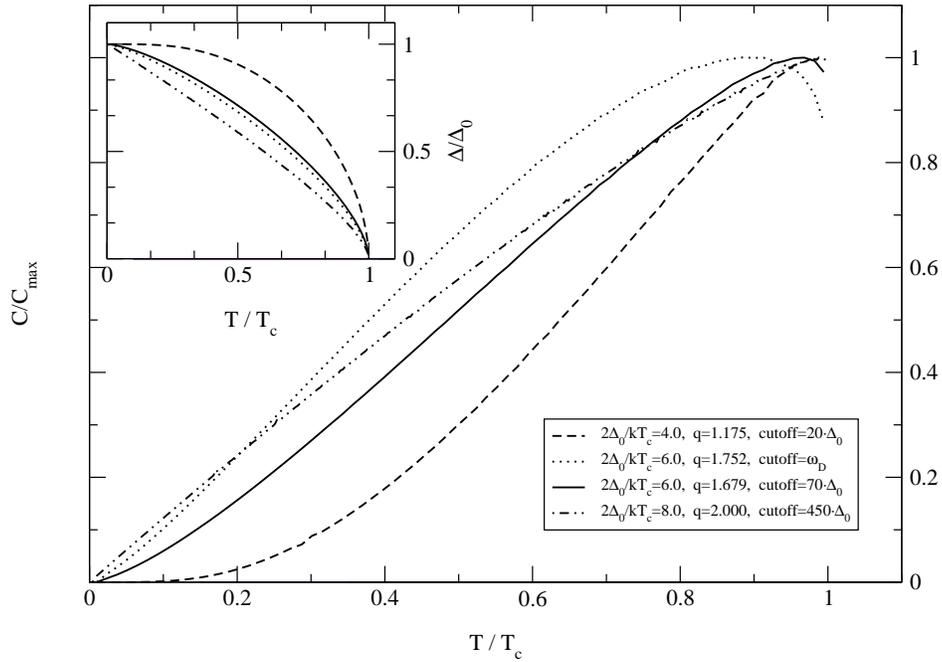}
\end{center}
\label{fig2}\caption{The normalized electronic specific heats $\frac{C}{C_{max}}$ vs. $\frac{T}{T_C}$ for various choices of $\frac{2\Delta_0}{k_B T_c}$, q and cutoffs where in each case $C_{max}$  corresponds to the maximum value of C. The corresponding values of the normalized gaps are shown in the inset. 
 }
\end{figure}

\begin{figure}[here]
\centering
\begin{center}
\includegraphics[scale=0.5,angle=270]{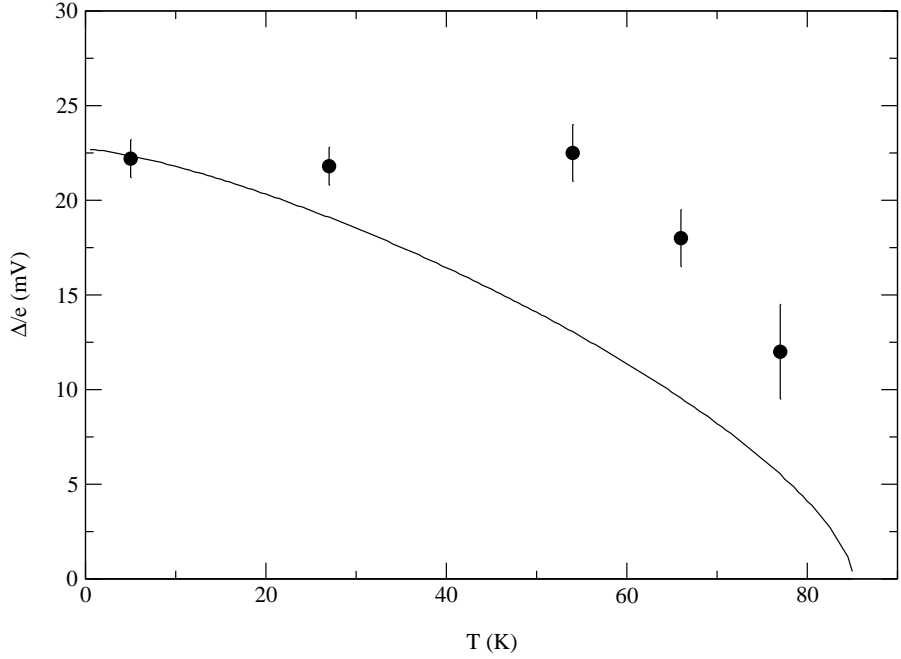}
\end{center}
\label{fig3}\caption{ Comparison of the temperature dependence of experimental values of $\Delta$ from Ref~\protect~\cite{briceno89}  for Bi$_2$Sr$_2$CaCu$_2$O$_8$ with  $T_c = 85K$ and  $\frac{2\Delta_0}{k_B T_c} = 6.2$  with the theoretical results for 
$q$=1.71 }
\end{figure}

The electronic specific heat capacity may  be expressed as
\begin{equation}
\label{thermsph}
C = T \frac{dS}{dT}
\end{equation}
where the  entropy is given by

\begin{equation}
\label{qSsystem}
S_q = 2k_B[\frac{\int d\varepsilon \{ f(E)-f(E)^q \}}{q-1} + \frac{\int d\varepsilon \{(1-f(E))-(1-f(E))^q\}}{q-1}]
\end{equation}
and we
use $N(0) = \frac{m}{2\pi\hbar^2}$ for the density of states in 2D and the bare electron mass 
$m_e$ in all cases. The effect of the cutoff  on the specific heat
is shown in Figure 2 for LSCO ($T_c = 36K$).  
The linear nature of the specific heat can be seen over most of the superconducting region with $\hbar\omega_D = 390K $ as the cutoff. Clearly the situation deteriotes as the critical temperature is reached. Increasing the cutoff to $70 \Delta_0$ yields a more linear form of the specific heat  near $T_c$
with little change in the shape of the gap (see inset in Fig. 2). Changing the cutoff from $\hbar\omega_D$ to $70 \Delta_0$ increases the $\gamma$ from 5.5 mJ/mol$\cdot$K$^2$ to 10.7 mJ/mol$\cdot$K$^2$ with a slight change in q (from 1.752 to 1.679).
Note that in spite of the fact that the slopes of the specific heat may be altered by using an effective mass $(\gamma \to \frac{m_e^*}{m_e}\gamma)$,
these values are not in disagreement with  many of the experimental results which seem to lie between 3 and 12 mJ/mol$\cdot$K$^2$ for most cuprates 
\cite{ginsberg90} For comparison  the ratios of $\frac{2\Delta_0}{k_B T_c} = 4.0$ and $8.0$, 
the corresponding results for the specific heats and the gaps are shown.
Note the specific heat is much more linear when
larger cutoffs are used and that the nonlinearity around $T_c$  disappears almost completely  for 
$\frac{2\Delta_0}{k_B T_c} = 8.0$. In this case $\gamma$ = 20.0 J/mol$\cdot$K$^2$ .

In  Fig. 3 the experimental data obtained by Briceno and Zettl \cite{briceno89} for the temperature dependence of the
gap of Bi$_2$Sr$_2$CaCu$_2$O$_8$ with  $T_c = 85K$ and  $\frac{2\Delta_0}{k_B T_c} = 6.2$ are compared with the theoretical results with 
$q$=1.717. In spite of a linear specific heat ($\gamma$ = 13.9 mJ/mol$\cdot$K$^2$ using $m_e$) which is quite consistent with experiment 
\cite{ginsberg90}, the shape of the gap differs slightly from that obtained experimentally.

 Due to the type II nature of the cuprate oxides, a full analysis of the critical magnetic field would require including contributions to
the free energy resulting from the flux lattice as well as the mutual energy of the surface current and flux line currents.  We therefore 
present only a qualitative analysis of the thermodynamic critical field, $B_c$, and make no attempt at solving either the lower- 
$(B_{c1})$ or the upper- $(B_{c2})$ critical fields.
  
\begin{figure}[h]
\centering
\begin{center}
\includegraphics[scale=0.5,angle=270]{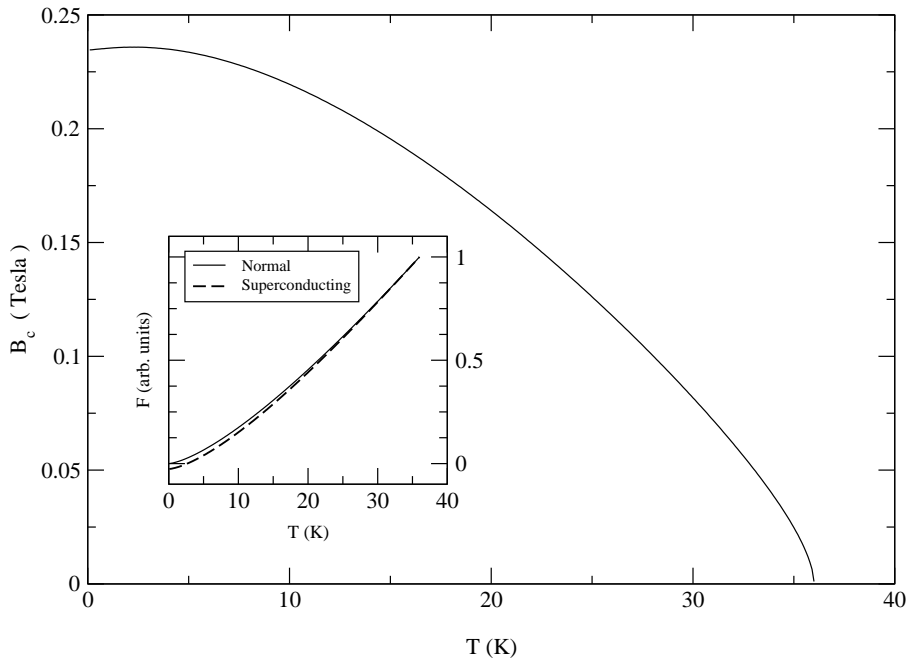}
\end{center}
\label{fig4}\caption{ The critical magnetic field $B_c$ vs. T obtained from free energies given in the inset.}
\end{figure}

In 2D,  the thermodynamic critical field is given by

\begin{equation}
\label{critfield1}
\frac{B_c^2}{2\mu_0} = F_n - F_s
\end{equation}
where $F_n$ and $F_s$ are the free energies in the normal- and superconducting states respectively and $\mu_0$ is the permeability of 
free space.  The free energy in the superconducting state is given by

\begin{equation}
F_s = N(0)(\hbar\omega)^2[[1 +(\frac{\Delta_0}{\hbar\omega})^2]^\frac{1}{2}-1] 
-2N(0)\int d\varepsilon(\frac{2\varepsilon^2 +\Delta^2}{E})f_q(\beta E)
\end{equation}
\begin{equation}
\label{Fsuper}
-\frac{4kT}{q-1}N(0)\int d\varepsilon[(1-f_q(\beta E))^q - (1-f_q(\beta E))].
\end{equation}
$F_n$ is obtained by setting $\Delta=0$ and using the value of q determined from solution of the corresponding gap equation. This implies that in our generalized BCS model the normal state is also described by the generalized statistics of Tsallis since $B_c$ must vanish at 
$T_C$.
Figure 4 shows the  thermodynamic field obtained for a superconductor with $\frac{2\Delta_0}{k_BT_c} = 6.0$ and $T_c$ = 36K .  The inset shows the behaviour of the free energies.  They differ  from the weak coupled normal superconductor case  where  the free energy is always negative except 
at $T=0$ where it vanishes.   The difference, however, remains a positive definite quantitiy and no problems arise in calculating $B_c$.  
The zero temperature result of figure 4 agrees exceptionally well with an extrapolation based on experimental results due to \cite{li93}. 
They find, for a La$_{1.846}$Sr$_{0.154}$CuO$_4$ crystal with $T_c$ = 35K that $B_c$ = 0.251 T, while our result is 0.234 T.

\begin{figure}[h]
\centering
\begin{center}
\includegraphics[scale=0.5,angle=270]{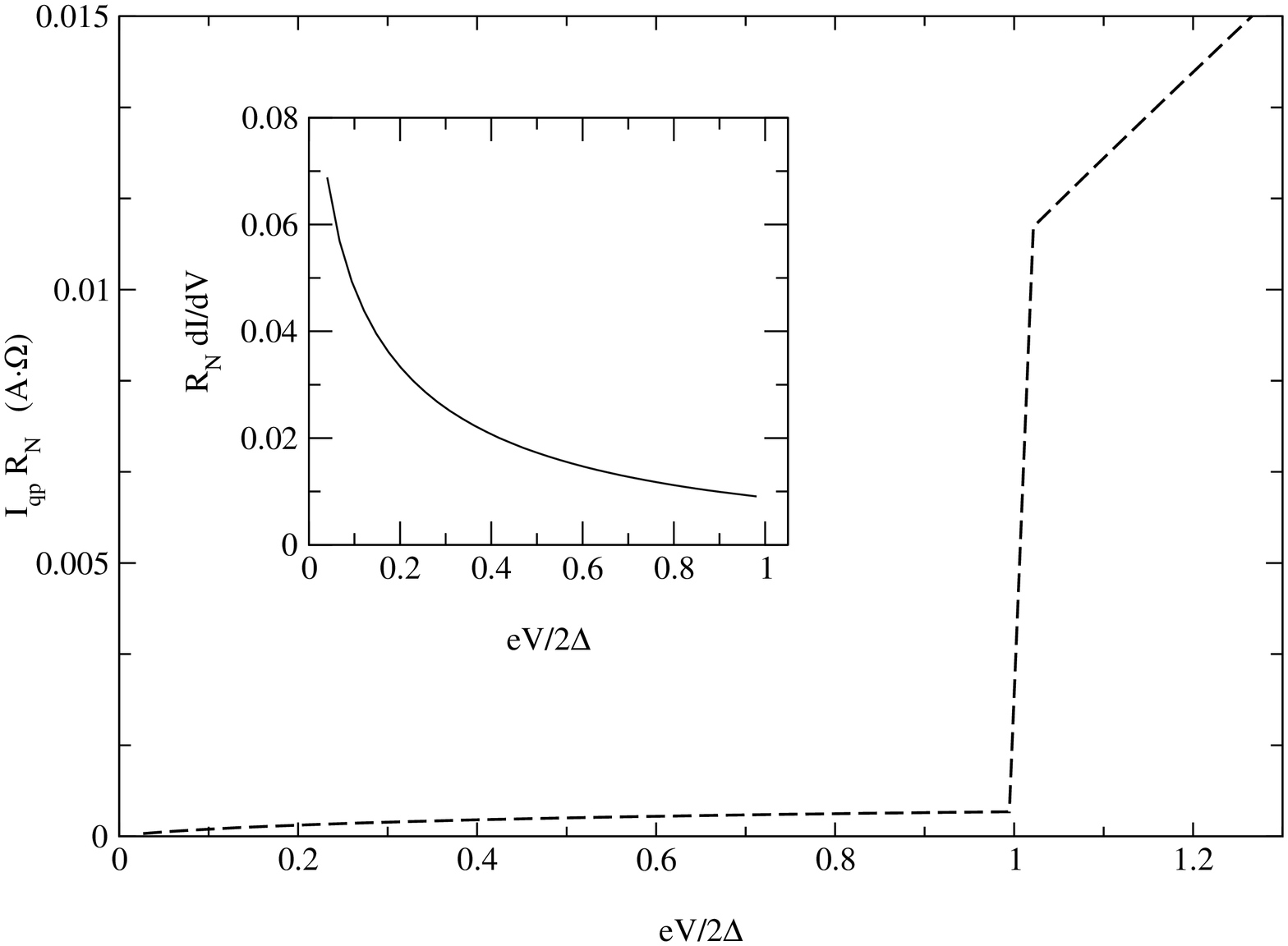}
\end{center}
\label{fig5}\caption{$\frac{dI}{dV} $vs.$\frac{eV}{2\Delta} $ (inset) obtained from the quasi-particle contribution to the Josephson tunneling current in the region eV$<$ 2$\Delta$, as shown in the main figure.   }
\end{figure}

The Josephson quasi-particle current was first treated in detail by Shapiro \textit{et al.} \cite{shapiro62} (the other terms are considered elsewhere \cite{harris74}).  The quasi-particle current between two different superconductors is given by the
integral:
\begin{equation}
\label{Iqp}
I_{qp}=\frac{1}{eR_N} \int^\infty_{-\infty}[f(\omega)-f(\omega+eV)]|\omega||\omega+eV| \frac{\theta(|\omega|-\Delta_1) \theta(|\omega + eV|-\Delta_2)}{(\omega^2 - \Delta^2_1)^\frac{1}{2}[(\omega+eV)^2-\Delta^2_2]^\frac{1}{2}}d\omega
\end{equation}
where V is the applied voltage,  e the electron charge, R$_N$ the junction resistance and $\Delta_1$, $\Delta_2$ the energy gap on either 
side of the tunneling junction.  In Figure 5 we show the result of evaluating equation (\ref{Iqp}) at 5K for a junction with identical 
superconductors having $\frac{2\Delta_0}{k_B T_c}=6.0$ and using the generalized distribution function with $q$=1.679.  Of interest is the fact that $I_{qp}$ is 
not negligible below an applied voltage of $2\Delta/$e.  The equivalent calculation using BG statistics yields a 
vanishing contribution in this region.     
Included in the figure is the derivative $\frac{dI}{dV}$.  Although we do not obtain all of the structure observed  experimentally 
in LSCO\cite{alff98}, the existence of a $I_{qp}$ contribution raises the possibility that interference due to other admixtures, perhaps d-wave or Andreev bound states, will produce the sought after structure.  Analysis of the other contributions using generalized statistics does not 
appreciably alter the results in the  eV $< 2\Delta$ region.  The use of generalized statistics,
in the BCS equations, however, allows the experimentally observed
discontinuity in the quasi-particle tunneling current to be interpreted as being coincident with the energy gap at low T, whilst 
consistently predicting the gap to vanish at the correct critical temperature.

In conclusion we have shown that a simple 2D s-wave BCS pairing model which incorporates generalized statistics can provide an adequate description of many of the main features of  high temperature superconductivity.
Non-phonon mediated interactions, however, appear to play a role as the critical temperature increases.

\listoffigures

\end{document}